\documentclass[runningheads]{llncs}
\usepackage[ruled,vlined]{algorithm2e}
\usepackage{amssymb}
\usepackage{mathtools}
\usepackage[capitalise]{cleveref}
\usepackage{tikz}
\usepackage{mathrsfs}
\usepackage[nounderscore]{syntax}
\usepackage{blkarray}
\usepackage[binary-units]{siunitx}
\usepackage[inline,shortlabels]{enumitem}
\usepackage{capt-of}
\usepackage[caption=false]{subfig}
\usepackage{booktabs}
\usepackage[misc,geometry]{ifsym}

\usetikzlibrary{arrows.meta}
\spnewtheorem{constraint}{Constraint}{\bfseries}{\itshape}

\renewcommand\fbox{\fcolorbox{red}{white}}
\newcommand{\hilight}[1]{\setlength{\fboxsep}{1pt}\colorbox{lightgray}{#1}}
\newcommand{\hlitem}{\stepcounter{enumi}\item[\hilight{\theenumi}]}

\newcommand{\logical}[1]{{\normalfont \texttt{#1}}}
\newcommand{\variable}[1]{\texttt{\textup{#1}}}
\newcommand{\arrayd}[3]{\variable{{#1}[}{#2}\variable{]} \in {#3}}
\newcommand{\arrayt}[3]{\variable{{#3}} : \variable{{#1}[}{#2}\variable{]}}

\newcommand{\predicates}{\mathcal{P}}
\newcommand{\variables}{\mathcal{V}}
\newcommand{\constants}{\mathcal{C}}
\newcommand{\tokens}{\mathcal{T}}
\newcommand{\arities}{\mathcal{A}}
\newcommand{\maxArity}{\mathcal{M}_{\mathcal{A}}}
\newcommand{\maxNumNodes}{\mathcal{M}_{\mathcal{N}}}
\newcommand{\maxNumClauses}{\mathcal{M}_{\mathcal{C}}}

\DeclareMathOperator{\Determined}{\Delta}
\DeclareMathOperator{\Undetermined}{\Upsilon}
\DeclareMathOperator{\AlmostDetermined}{\Gamma}

\Crefname{constraint}{Constraint}{Constraints}
\crefname{section}{Sect.}{Sects.}

\begin{document}

\title{Generating Random Logic Programs Using Constraint Programming}
\author{Paulius Dilkas\inst{1} \and Vaishak Belle\inst{1,2}$^{(\textrm{\Letter})}$}
\authorrunning{P. Dilkas and V. Belle}
\institute{University of Edinburgh, Edinburgh, UK \\
  \email{p.dilkas@sms.ed.ac.uk, vaishak@ed.ac.uk} \and Alan Turing Institute,
  London, UK}

\maketitle

\begin{abstract}
  Testing algorithms across a wide range of problem instances is crucial to
  ensure the validity of any claim about one algorithm's superiority over
  another. However, when it comes to inference algorithms for probabilistic
  logic programs, experimental evaluations are limited to only a few programs.
  Existing methods to generate random logic programs are limited to
  propositional programs and often impose stringent syntactic restrictions. We
  present a novel approach to generating random logic programs and random
  probabilistic logic programs using constraint programming, introducing a new
  constraint to control the independence structure of the underlying probability
  distribution. We also provide a combinatorial argument for the correctness of
  the model, show how the model scales with parameter values, and use the model
  to compare probabilistic inference algorithms across a range of synthetic
  problems. Our model allows inference algorithm developers to evaluate and
  compare the algorithms across a wide range of instances, providing a detailed
  picture of their (comparative) strengths and weaknesses.

  \keywords{Constraint programming \and Probabilistic logic programming \and
    Statistical relational learning}
\end{abstract}

\section{Introduction}

Unifying logic and probability is a long-standing challenge in artificial
intelligence \cite{DBLP:journals/cacm/Russell15}, and, in that regard,
statistical relational learning (SRL) has developed into an exciting area that
mixes machine learning and symbolic (logical and relational) structures. In
particular, probabilistic logic programs---including languages such as
\textsc{PRISM} \cite{DBLP:conf/ijcai/SatoK97}, \textsc{ICL}
\cite{DBLP:journals/ai/Poole97}, and \textsc{ProbLog}
\cite{DBLP:conf/ijcai/RaedtKT07}---are promising frameworks for codifying
complex SRL models. With the enhanced structure, however, inference becomes more
challenging. At the moment, we have no precise way of evaluating and comparing
inference algorithms. Incidentally, if one were to survey the literature, one
often finds that an inference algorithm is only tested on a small number (1--4)
of data sets
\cite{DBLP:conf/ecai/BruynoogheMKGVJR10,DBLP:journals/tplp/KimmigDRCR11,DBLP:conf/ijcai/VlasselaerBKMR15},
originating from areas such as social networks, citation patterns, and
biological data. But how confident can we be that an algorithm works well if it
is only tested on a few problems?

About thirty years ago, SAT solving technology was dealing with a similar lack
of clarity \cite{DBLP:journals/ai/SelmanML96}. This changed with the study of
generating random SAT instances against different input parameters (e.g.,
clause length and the total number of variables) to better understand the
behaviour of algorithms and their ability to solve random synthetic problems.
Unfortunately, when it comes to generating random logic programs, all approaches
so far focused exclusively on propositional programs
\cite{DBLP:conf/ijcai/AmendolaRT17,DBLP:journals/ai/AmendolaRT20,DBLP:journals/tplp/WangWM15,DBLP:conf/iclp/ZhaoL03},
often with severely limiting conditions such as two-literal clauses
\cite{DBLP:conf/iclp/Namasivayam09,DBLP:conf/lpnmr/NamasivayamT09} or clauses of
the form $a \gets \neg b$ \cite{DBLP:journals/tocl/WenWSL16}.

In this work (\cref{sec:heads,sec:bodies,sec:variable_symmetry}), we introduce a
constraint-based representation for logic programs based on simple parameters
that describe the program's size, what predicates and constants it uses, etc.
This representation takes the form of a \emph{constraint satisfaction problem}
(CSP), i.e., a set of discrete variables and restrictions on what values they
can take. Every solution to this problem (as output by a constraint solver)
directly translates into a logic program. One can either output all
(sufficiently small) programs that satisfy the given conditions or use random
value-ordering heuristics and restarts to generate random programs. For sampling
from a uniform distribution, the CSP can be transformed into a belief network
\cite{DBLP:conf/aaai/DechterKBE02}. In fact, the same model can generate both
probabilistic programs in the syntax of \textsc{ProbLog}
\cite{DBLP:conf/ijcai/RaedtKT07} and non-probabilistic \textsc{Prolog} programs.
To the best of our knowledge, this is the first work that
\begin{enumerate*}[(a)]
\item addresses the problem of generating random logic programs in its full
  generality (i.e., including first-order clauses with variables), and
\item compares and evaluates inference algorithms for probabilistic logic
  programs on more than a handful of instances.
\end{enumerate*}

A major advantage of a constraint-based approach is the ability to add
additional constraints as needed, and to do that efficiently (compared to
generate-and-test approaches). As an example of this, in \cref{sec:independence}
we develop a custom constraint that, given two predicates $\mathsf{P}$ and
$\mathsf{Q}$, ensures that any ground atom with predicate $\mathsf{P}$ is
independent of any ground atom with predicate $\mathsf{Q}$. In this way, we can
easily regulate the independence structure of the underlying probability
distribution. In \cref{sec:counting} we also present a combinatorial argument
for correctness that counts the number of programs that the model produces for
various parameter values. We end the paper with two experimental results in
\cref{sec:experiments}: one investigating how the constraint model scales when
tasked with producing more complex programs, and one showing how the model can
be used to evaluate and compare probabilistic inference algorithms.

Overall, our main contributions are concerned with logic programming-based
languages and frameworks, which capture a major fragment of SRL
\cite{DBLP:series/synthesis/2016Raedt}. However, since probabilistic logic
programming languages are closely related to other areas of machine learning,
including (imperative) probabilistic programming
\cite{DBLP:journals/ml/RaedtK15}, our results can lay the foundations for
exploring broader questions on generating models and testing algorithms in
machine learning.

\section{Preliminaries}

The basic primitives of logic programs are \emph{constants}, \emph{(logic)
  variables}, and \emph{predicates} with their \emph{arities}. A \emph{term} is
either a variable or a constant, and an \emph{atom} is a predicate of arity $n$
applied to $n$ terms. A \emph{formula} is any well-formed expression that
connects atoms using conjunction $\land$, disjunction $\lor$, and negation
$\neg$. A \emph{clause} is a pair of a \emph{head} (which is an atom) and a
\emph{body} (which is a formula\footnote{Our model supports arbitrarily complex
  bodies of clauses (e.g., $\neg\mathsf{P}(X) \lor (\mathsf{Q}(X) \land
  \mathsf{P}(X))$) because \textsc{ProbLog} does too. However, one can easily
  restrict our representation of a body to a single conjunction of literals
  (e.g., $\mathsf{Q}(X) \land \neg\mathsf{P}(X))$) by adding a couple of
  additional constraints.}). A \emph{(logic) program} is a set of clauses, and a
\emph{\textsc{ProbLog} program} is a set of clause-probability pairs
\cite{DBLP:journals/tplp/FierensBRSGTJR15}.

In the world of CSPs, we also have \emph{(constraint) variables}, each with a
\emph{domain}, whose values are restricted using \emph{constraints}. All
constraint variables in the model are integer or set variables, however, if an
integer refers to a logical construct (e.g., a logical variable or a constant),
we will make no distinction between the two. We say that a constraint variable
is \emph{(fully) determined} if its domain (at the time) has exactly one value.
We let $\Box$ denote the absent/disabled value of an optional variable
\cite{DBLP:conf/cpaior/MearsSSTMW14}. We write $\arrayd{a}{b}{c}$ to mean that
$\variable{a}$ is an array of variables of length $b$ such that each element of
$\variable{a}$ has domain $c$. Similarly, we write $\arrayt{a}{b}{c}$ to denote
an array $\variable{a}$ of length $b$ such that each element of $\variable{a}$
has type $\variable{c}$. Finally, we assume that all arrays start with index
zero.

\paragraph{Parameters of the model.} We begin by defining sets and lists of the
primitives used in constructing logic programs: a list of predicates
$\predicates{}$, a list of their corresponding arities $\arities{}$ (so
$|\arities{}|$ = $|\predicates{}|$), a set of variables $\variables{}$, and a
set of constants $\constants{}$. Either $\variables{}$ or $\constants{}$ can be
empty, but we assume that $|\constants{}| + |\variables{}| > 0$. Similarly, the
model supports zero-arity predicates but requires at least one predicate to have
non-zero arity. For notational convenience, we also set $\maxArity{} = \max
\arities{}$. Next, we need a measure of how complex a body of a clause can be.
As we represent each body by a tree (see \cref{sec:bodies}), we set
$\maxNumNodes{} \ge 1$ to be the maximum number of nodes in the tree
representation of any clause. We also set $\maxNumClauses{}$ to be the maximum
number of clauses in a program. We must have that $\maxNumClauses{} \ge
|\predicates{}|$ because we require each predicate to have at least one clause
that defines it. The model supports enforcing predicate independence (see
\cref{sec:independence}), so a set of independent pairs of predicates is another
parameter. Since this model can generate probabilistic as well as
non-probabilistic programs, each clause is paired with a probability which is
randomly selected from a given list---our last parameter. For generating
non-probabilistic programs, one can set this list to $[1]$. Finally, we define
$\tokens{} = \{ \neg, \land, \lor, \top \}$ as the set of tokens that (together
with atoms) form a clause. All decision variables of the model can now be
divided into $2 \times \maxNumClauses{}$ separate groups, treating the body and
the head of each clause separately. We say that the variables are contained in
two arrays: $\arrayt{bodies}{\maxNumClauses{}}{Body}$ and
$\arrayt{heads}{\maxNumClauses{}}{Head}$.

\section{Heads of Clauses} \label{sec:heads}

We define the \emph{head} of a clause as a $\variable{predicate} \in \predicates{}
\cup \{ \Box \}$ and $\arrayd{arguments}{\maxArity{}}{\constants{} \cup
  \variables{}} \cup \{ \Box \}$. Here, we use $\Box$ to denote either a
disabled clause that we choose not to use or disabled arguments if the arity of
the $\variable{predicate}$ is less than $\maxArity{}$. The reason why we need a
separate value for the latter (i.e., why it is not enough to fix disabled
arguments to a single already-existing value) will become clear in
\cref{sec:variable_symmetry}. This $\variable{predicate}$ variable has a
corresponding arity that depends on the $\variable{predicate}$. We can
define $\variable{arity} \in [0, \maxArity{}]$ as the arity of the
$\variable{predicate}$ if $\variable{predicate} \in \predicates{}$ and zero
otherwise using the table constraint \cite{DBLP:conf/cpaior/MairyDL15}. This
constraint uses a set of pairs of the form $(p, a)$, where $p$ ranges over all
possible values of the $\variable{predicate}$, and $a$ is either the arity of
predicate $p$ or zero. Having defined arity, we can now fix the superfluous
arguments.
\begin{constraint} \label{constr:arity}
  For $i = 0, \dots, \maxArity{} - 1$, $\variable{arguments}[i] = \Box \iff i
  \ge \variable{arity}$.
\end{constraint}
We also add a constraint that each predicate should get at least one clause.
\begin{constraint}
  Let $P = \{ h.\variable{predicate} \mid h \in \variable{heads} \}$ be a
  multiset. Then
  \[
    \variable{nValues}(P) = \begin{cases}
      |\predicates{}| & \text{if } \variable{count}(\Box, P) = 0 \\
      |\predicates{}| + 1 & \text{otherwise,}
    \end{cases}
  \]
  where $\variable{nValues}(P)$ counts the number of unique values in $P$, and
  $\variable{count}(\Box, P)$ counts how many times $\Box$ appears in $P$.
\end{constraint}
Finally, we want to disable duplicate clauses but with one exception: there may
be more than one disabled clause, i.e., a clause with head $\variable{predicate}
= \Box$. Assuming a lexicographic order over entire clauses such that $\Box >
\mathsf{P}$ for all $\mathsf{P} \in \predicates{}$ and the head predicate is the
`first digit' of this representation, the following constraint disables
duplicates as well as orders the clauses.
\begin{constraint}
  For $i = 1, \dots, \maxNumClauses{} - 1$, if $\variable{heads}[i].\variable{predicate}
  \ne \Box$, then
  \[
    (\variable{heads}[i-1], \variable{bodies}[i-1]) < (\variable{heads}[i],
    \variable{bodies}[i]).
  \]
\end{constraint}

\section{Bodies of Clauses} \label{sec:bodies}

As was briefly mentioned before, the \emph{body} of a clause is represented by a
tree. It has two parts. First, there is the
$\arrayd{structure}{\maxNumNodes{}}{[0, \maxNumNodes{} - 1]}$ array that encodes
the structure of the tree using the following two rules:
$\variable{structure}[i] = i$ means that the $i$th node is a root, and
$\variable{structure}[i] = j$ (for $j \ne i$) means that the $i$th node's
parent is node $j$. The second part is the array
$\arrayt{values}{\maxNumNodes{}}{Node}$ such that $\variable{values}[i]$ holds
the value of the $i$th node, i.e., a representation of the atom or logical
operator.

We can use the $\variable{tree}$ constraint \cite{DBLP:conf/cp/FagesL11} to
forbid cycles in the $\variable{structure}$ array and simultaneously define
$\variable{numTrees} \in \{ 1, \dots, \maxNumNodes{} \}$ to count the number of
trees. We will view the tree rooted at the zeroth node as the main tree and
restrict all other trees to single nodes. For this to work, we need to make sure
that the zeroth node is indeed a root, i.e., fix $\variable{structure}[0] = 0$.
For convenience, we also define $\variable{numNodes} \in \{ 1, \dots,
\maxNumNodes{} \}$ to count the number of nodes in the main tree. We define it
as $\variable{numNodes} = \maxNumNodes{} - \variable{numTrees} + 1$.

\begin{example} \label{example:formula}
  Let $\maxNumNodes{} = 8$. Then $\neg\mathsf{P}(X) \lor (\mathsf{Q}(X) \land
  \mathsf{P}(X))$ can be encoded as:
  \begin{alignat*}{9}
    \variable{structure} &= [0, &&0, &&0, &&1, &&2, &&2, &&6, &&7], \quad
    &&\variable{numNodes} = 6, \\
    \variable{values} &= [{\lor}, &&{\neg}, &&{\land}, \mathsf{P}(&&X),
    \mathsf{Q}(&&X), \mathsf{P}(&&X), &&\top, &&\top], \quad
    &&\variable{numTrees} = 3.
  \end{alignat*}
\end{example}

Here, $\top$ is the value we use for the remaining one-node trees. The
elements of the $\variable{values}$ array are nodes. A \emph{node} has a
$\variable{name} \in \tokens{} \cup \predicates{}$ and
$\arrayd{arguments}{\maxArity{}}{\variables{} \cup \constants{} \cup \{ \Box
  \}}$. The node's $\variable{arity}$ can then be defined in the same way as in
\cref{sec:heads}. Furthermore, we can use \cref{constr:arity} to again disable
the extra arguments.

\begin{example}
  Let $\maxArity{} = 2$, $X \in \variables{}$, and let $\mathsf{P}$ be a
  predicate with arity 1. Then the node representing atom $\mathsf{P}(X)$ has:
    $\variable{name} = \mathsf{P}$, $\variable{arguments} = [X, \Box]$,
    $\variable{arity} = 1$.
\end{example}

We need to constrain the forest represented by the $\variable{structure}$
array together with its $\variable{values}$ to eliminate symmetries and adhere
to our desired format. First, we can recognise that the order of the elements in
the $\variable{structure}$ array does not matter, i.e., the structure is only
defined by how the elements link to each other, so we can add a constraint for
sorting the $\variable{structure}$ array. Next, since we already have a
variable that counts the number of nodes in the main tree, we can fix the
structure and the values of the remaining trees to some constant values.
\begin{constraint}
  For $i = 1, \dots, \maxNumNodes{} - 1$, if $i < \variable{numNodes}$, then
  \[
    \variable{structure}[i] = i, \quad \text{and} \quad
    \variable{values}[i].\variable{name} = \top,
  \]
  else $\variable{structure}[i] < i$.
\end{constraint}
The second part of this constraint states that every node in the main tree
except the zeroth node cannot be a root and must have its parent located to
the left of itself. Next, we classify all nodes into three classes: predicate
(or empty) nodes, negation nodes, and conjunction/disjunction nodes based on the
number of children (zero, one, and two, respectively).
\begin{constraint} \label{constraint:node_types}
  For $i = 0, \dots, \maxNumNodes{} - 1$, let $C_i$ be the number of times $i$
  appears in the \variable{structure} array with index greater than $i$. Then
  \begin{align*}
    C_i = 0 &\iff \variable{values}[i].\variable{name} \in \predicates{} \cup \{ \top \},\\
    C_i = 1 &\iff \variable{values}[i].\variable{name} = \neg,\\
    C_i > 1 &\iff \variable{values}[i].\variable{name} \in \{ \land, \lor \}.
  \end{align*}
\end{constraint}
The value $\top$ serves a twofold purpose: it is used as the fixed value for
nodes outside the main tree, and, when located at the zeroth node, it can
represent a clause with an empty body. Thus, we can say that only root nodes can
have $\top$ as the value.
\begin{constraint}
  For $i = 0, \dots, \maxNumNodes{} - 1$,
  \[
    \variable{structure}[i] \ne i \implies
    \variable{values}[i].\variable{name} \ne \top.
  \]
\end{constraint}
Finally, we add a way to disable a clause by setting its head predicate to
$\Box$.
\begin{constraint}
  For $i = 0, \dots, \maxNumClauses{} - 1$, if
  $\variable{heads}[i].\variable{predicate} = \Box$, then
  \[
    \variable{bodies}[i].\variable{numNodes} = 1, \quad \text{and}
    \quad \variable{bodies}[i].\variable{values}[0].\variable{name} = \top.
  \]
\end{constraint}

\section{Variable Symmetry Breaking} \label{sec:variable_symmetry}

Ideally, we want to avoid generating programs that are equivalent in the sense
that they produce the same answers to all queries. Even more importantly, we
want to avoid generating multiple internal representations that ultimately
result in the same program. This is the purpose of \emph{symmetry-breaking
  constraints}, another important benefit of which is that the constraint
solving task becomes easier \cite{DBLP:conf/cp/Walsh06}. Given any clause, we
can permute the variables in that clause without changing the meaning of the
clause or the entire program. Thus, we want to fix the order of variables.
Informally, we can say that variable $X$ goes before variable $Y$ if the first
occurrence of $X$ in either the head or the body of the clause is before the
first occurrence of $Y$. Note that the constraints described in this section
only make sense if $|\variables{}| > 1$ and that all definitions and constraints
here are on a per-clause basis.
\begin{definition}
  Let $N = \maxArity{} \times (\maxNumNodes{} + 1)$, and let
  $\variable{terms}[N] \in \constants{} \cup \variables{} \cup \{ \Box
  \}$ be a flattened array of all arguments in a particular clause. Then we can
  use a channeling constraint to define $\variable{occ}[|\constants{}| +
  |\variables{}| + 1]$ as an array of subsets of $\{ 0, \dots, N-1 \}$ such that
  for all $i = 0, \dots, N - 1$, and $t \in \constants{} \cup \variables{} \cup
  \{ \Box \}$,
  \[
    i \in \variable{occ}[t] \quad \iff \quad
    \variable{terms}[i] = t.
  \]
\end{definition}
Next, we introduce an array that holds the first occurrence of each variable.
\begin{definition}
  Let $\arrayd{intros}{|\variables{}|}{\{ 0, \dots, N \}}$ be such that
  for $v \in \variables{}$,
  \[
    \variable{intros}[v] = \begin{cases}
      1 + \min \variable{occ}[v] & \text{if }
      \variable{occ}[v] \ne \emptyset\\
      0 & \text{otherwise.}
    \end{cases}
  \]
\end{definition}
Here, a value of zero means that the variable does not occur in the clause (this
choice is motivated by subsequent constraints). As a consequence, all other
indices are shifted by one. Having set this up, we can now eliminate variable
symmetries simply by sorting $\variable{intros}$. In other words, we constrain
the model so that the variable listed first (in whatever order $\variables{}$ is
presented in) has to occur first in our representation of a clause.

\begin{example} \label{example:sibling}
  Let $\constants{} = \emptyset$, $\variables{} = \{ X, Y, Z \}$, $\maxArity{} =
  2$, $\maxNumNodes{} = 3$, and consider the clause $\mathsf{sibling}(X, Y)
  \gets \mathsf{parent}(X, Z) \land \mathsf{parent}(Y, Z)$. Then
  \begin{align*}
    \variable{terms} &= [X, Y, \Box, \Box, X, Z, Y, Z], \\
    \variable{occ} &= [\{ 0, 4 \}, \{ 1, 6 \}, \{ 5, 7 \}, \{ 2, 3 \}], \\
    \variable{intros} &= [0, 1, 5],
  \end{align*}
  where the $\Box$'s correspond to the conjunction node.
\end{example}

We end the section with several redundant constraints that make the CSP easier
to solve. First, we can state that the positions occupied by different terms
must be different.
\begin{constraint} \label{constraint:all_diff}
  For $u \ne v \in \constants{} \cup \variables{} \cup \{ \Box \}$,
  $\variable{occ}[u] \cap \variable{occ}[v] = \emptyset$.
\end{constraint}
The reason why we use zero to represent an unused variable is so that we could
now use the `all different except zero' constraint for the $\variable{intros}$
array. We can also add another link between $\variable{intros}$ and
$\variable{occ}$ that essentially says that the smallest element of a set is an
element of the set.
\begin{constraint}
  For $v \in \variables{}$, $\variable{intros}[v] \ne 0 \iff
  \variable{intros}[v] - 1 \in \variable{occ}[v]$.
\end{constraint}
Finally, we define an auxiliary set variable to act as a set of possible values
that $\variable{intros}$ can take. Let $\variable{potentials} \subseteq \{ 0,
\dots, N \}$ be such that for $v \in \variables{}$, $\variable{intros}[v] \in
\variable{potentials}$. Using this new variable, we can add a constraint saying
that non-predicate nodes in the tree representation of a clause cannot have
variables as arguments.
\begin{constraint} \label{constraint:potentialIntroductions}
  For $i = 0, \dots, \maxNumNodes{} - 1$, let
  \[
    S = \{ \maxArity{} \times (i + 1) + j + 1 \mid j = 0, \dots, \maxArity{} - 1
    \}.
  \]
  If $\variable{values}[i].\variable{name} \not\in \predicates{}$, then
  $\variable{potentials} \cap S = \emptyset$.
\end{constraint}

\section{Counting Programs} \label{sec:counting}

To demonstrate the correctness of the model, this section derives combinatorial
expressions for counting the number of programs with up to $\maxNumClauses{}$
clauses and up to $\maxNumNodes{}$ nodes per clause, and arbitrary
$\predicates{}$, $\arities{}$, $\variables{}$, and $\constants{}$. Being able to
establish two ways to generate the same sequence of numbers (i.e., numbers of
programs with certain properties and parameters) allows us to gain confidence
that the constraint model accurately matches our intentions. For this section,
we introduce the term \emph{total arity} of a body of a clause to refer to the
sum total of arities of all predicates in the body.

We will first consider clauses with \emph{gaps}, i.e., without taking variables
and constants into account. Let $T(n, a)$ denote the number of possible clause
bodies with $n$ nodes and total arity $a$. Then $T(1, a)$ is the number of
predicates in $\predicates{}$ with arity $a$, and the following recursive
definition can be applied for $n > 1$:
\[
  T(n, a) = T(n-1, a) + 2\sum_{\substack{c_1 + \dots + c_k = n - 1,\\ 2 \le k
      \le \frac{a}{\min \arities{}},\\ c_i \ge 1 \text{ for all } i}}
  \sum_{\substack{d_1 + \dots + d_k = a,\\ d_i \ge \min \arities{} \text{ for
        all } i}} \prod_{i=1}^k T(c_i, d_i).
\]
The first term here represents negation, i.e., negating a formula consumes
one node but otherwise leaves the task unchanged. If the first operation is not
a negation, then it must be either conjunction or disjunction (hence the
coefficient `2'). In the first sum, $k$ represents the number of children of the
root node, and each $c_i$ is the number of nodes dedicated to child $i$. Thus,
the first sum iterates over all possible ways to partition the remaining $n-1$
nodes. Similarly, the second sum considers every possible way to partition the
total arity $a$ across the $k$ children nodes. We can then count the number of
possible clause bodies with total arity $a$ (and any number of nodes) as
\[
  C(a) = \begin{cases}
    1 & \text{if } a = 0\\
    \sum_{n=1}^{\maxNumNodes{}} T(n, a) & \text{otherwise.}
  \end{cases}
\]

The number of ways to select $n$ terms is
\[
  P(n) = |\constants{}|^n + \sum_{\substack{1 \le k \le |\variables{}|, \\ 0 =
      s_0 < s_1 < \dots < s_k < s_{k+1} = n+1}} \prod_{i=0}^k (|\constants{}| +
  i)^{s_{i+1} - s_i - 1}.
\]
The first term is the number of ways to select $n$ constants. The parameter $k$
is the number of variables used in the clause, and $s_1, \dots, s_k$ mark the
first occurrence of each variable. For each gap between any two introductions
(or before the first introduction, or after the last introduction), we have
$s_{i+1}-s_i-1$ spaces to be filled with any of the $|\constants{}|$ constants
or any of the $i$ already-introduced variables.

Let us order the elements of $\predicates{}$, and let $a_i$ be the arity of the
$i$th predicate. The number of programs is then:
\begin{equation} \label{eq:counting}
  \sum_{\substack{ \sum_{i=1}^{|\predicates{}|} h_i = n,\\
      |\predicates{}| \le n \le \maxNumClauses{},\\
      h_i \ge 1 \text{ for all } i}} \prod_{i=1}^{|\predicates{}|}
  \binom{\sum_{a=0}^{\maxArity{} \times \maxNumNodes{}} C(a) P(a+a_i)}{h_i},
\end{equation}
Here, we sum over all ways to distribute $|\predicates{}| \le n \le
\maxNumClauses{}$ clauses among $|\predicates{}|$ predicates so that each
predicate gets at least one clause. For each predicate, we can then count the
number of ways to select its clauses out of all possible clauses. The number of
possible clauses can be computed by considering each possible arity $a$, and
multiplying the number of `unfinished' clauses $C(a)$ by the number of ways to
select the required $a+a_i$ terms in the body and the head of the clause.
Finally, we compare the numbers produced by \eqref{eq:counting} with the numbers
of programs generated by our model in \num{1032} different scenarios, thus
showing that the combinatorial description developed in this section matches the
model's behaviour.

\section{Stratification and Independence} \label{sec:independence}

\emph{Stratification} is a condition necessary for probabilistic logic programs
\cite{DBLP:conf/padl/MantadelisR17} and often enforced on logic programs
\cite{DBLP:journals/tcs/Bidoit91} that helps to ensure a unique answer to every
query. This is achieved by restricting the use of negation so that any program
$\mathscr{P}$ can be partitioned into a sequence of programs $\mathscr{P} =
\bigsqcup_{i=1}^n \mathscr{P}_i$ such that, for all $i$, the negative literals
in $\mathscr{P}_i$ can only refer to predicates defined in $\mathscr{P}_j$ for
$j \le i$ \cite{DBLP:journals/tcs/Bidoit91}.

\emph{Independence}, on the other hand, is defined on a pair of predicates (say,
$\mathsf{P}, \mathsf{Q} \in \mathcal{P}$) and can be interpreted in two ways.
First, if $\mathsf{P}$ and $\mathsf{Q}$ are independent, then any ground atom of
$\mathsf{P}$ is independent of any ground atom of $\mathsf{Q}$ in the underlying
probability distribution of the probabilistic program. Second, the part of the
program needed to fully define $\mathsf{P}$ is disjoint from the part of the
program needed to define $\mathsf{Q}$.

These two seemingly disparate concepts can be defined using the same building
block, i.e., a predicate dependency graph. Let $\mathscr{P}$ be a probabilistic
logic program with its set of predicates $\predicates{}$. Its \emph{(predicate)
  dependency graph} is a directed graph $G_{\mathscr{P}}$ with elements of
$\predicates{}$ as nodes and an edge between $\mathsf{P}, \mathsf{Q} \in
\predicates{}$ if there is a clause in $\mathscr{P}$ with $\mathsf{Q}$ as the
head and $\mathsf{P}$ mentioned in the body. We say that the edge is
\emph{negative} if there exists a clause with $\mathsf{Q}$ as the head and at
least one instance of $\mathsf{P}$ at the body such that the path from the root
to the $\mathsf{P}$ node in the tree representation of the clause passes through
at least one negation node; otherwise, it is \emph{positive}. We say that
$\mathscr{P}$ (or $G_{\mathscr{P}}$) has a \emph{negative cycle} if
$G_{\mathscr{P}}$ has a cycle with at least one negative edge. A program
$\mathscr{P}$ is \emph{stratified} if $G_{\mathscr{P}}$ has no negative
cycles.\footnote{This definition is an extension of a well-known result for
  logic programs \cite{DBLP:journals/jlp/BalbinPRM91} to probabilistic logic
  programs with arbitrary complex clause bodies.} Thus a simple entailment
algorithm for stratification can be constructed by selecting all clauses, all
predicates of which are fully determined, and looking for negative cycles in the
dependency graph constructed based on those clauses using an algorithm such as
Bellman-Ford.

For any predicate $\mathsf{P} \in \predicates{}$, the set of \emph{dependencies}
of $\mathsf{P}$ is the smallest set $D_{\mathsf{P}}$ such that $\mathsf{P} \in
D_{\mathsf{P}}$, and, for every $\mathsf{Q} \in D_{\mathsf{P}}$, all direct
predecessors of $\mathsf{Q}$ in $G_{\mathscr{P}}$ are in $D_{\mathsf{P}}$. Two
predicates $\mathsf{P}$ and $\mathsf{Q}$ are \emph{independent} if
$D_{\mathsf{P}} \cap D_{\mathsf{Q}} = \emptyset$.

\begin{example} \label{ex:program}
  Consider the following (fragment of a) program:
  \begin{align}
    \mathsf{sibling}(X, Y) &\gets \mathsf{parent}(X, Z) \land \mathsf{parent}(Y, Z), \nonumber \\
    \mathsf{father}(X, Y) &\gets \mathsf{parent}(X, Y) \land \neg\mathsf{mother}(X, Y). \label[clause]{eq:example_clause}
  \end{align}
  Its predicate dependency graph is in \cref{fig:predicate_dependencies}.
  Because of the negation in \eqref{eq:example_clause}, the edge from
  $\mathsf{mother}$ to $\mathsf{father}$ is negative, while the other two edges
  are positive. The dependencies of each predicate are:
  \begin{alignat*}{3}
    D_{\mathsf{parent}} &= \{ \mathsf{parent} \}, \quad && D_{\mathsf{sibling}}
    &&= \{\mathsf{sibling}, \mathsf{parent} \},\\
    D_{\mathsf{mother}} &= \{ \mathsf{mother} \}, \quad && D_{\mathsf{father}}
    &&= \{ \mathsf{father}, \mathsf{mother}, \mathsf{parent} \}.
  \end{alignat*}
  Hence, we have two pairs of independent predicates, i.e., $\mathsf{mother}$ is
  independent of $\mathsf{parent}$ and $\mathsf{sibling}$.
\end{example}

\begin{figure}[t]
  \centering
  \begin{minipage}{.49\textwidth}
    \centering
    \begin{tikzpicture}
      \node[draw] (parent) at (0, 0.5) {$\mathsf{parent}$};
      \node[draw] (mother) at (0, -0.5) {$\mathsf{mother}$};
      \node[draw] (sibling) at (2, 0.5) {$\mathsf{sibling}$};
      \node[draw] (father) at (2, -0.5) {$\mathsf{father}$};
      \draw[-{Stealth[scale=1.5]}] (parent) edge node[above] {$+$} (sibling);
      \draw[-{Stealth[scale=1.5]}] (parent) edge node[above] {$+$} (father);
      \draw[-{Stealth[scale=1.5]}] (mother) edge node[below] {$-$} (father);
    \end{tikzpicture}%
    \captionof{figure}{The predicate dependency graph of the program from
      \protect{\cref{ex:program}}. Positive edges are labelled with `$+$', and
      negative edges with `$-$'.}
    \label{fig:predicate_dependencies}
  \end{minipage}
  \begin{minipage}{.49\textwidth}
    \centering
    \captionof{table}{Types of (potential) dependencies of a predicate
      $\mathsf{P}$ based on the number of undetermined edges on the path from
      the dependency to $\mathsf{P}$}
    \label{tbl:dependencies}
    \begin{tabular}{clc}
      \toprule
      Edges & Name & Notation \\
      \midrule
      $0$ & Determined & $\Determined(p)$ \\
      $1$ & Almost determined & $\AlmostDetermined(p, s, t)$ \\
      $>1$ & Undetermined & $\Undetermined(p)$ \\
      \bottomrule
    \end{tabular}
  \end{minipage}
\end{figure}

Since the definition of independence relies on the dependency graph, we can
represent this graph as an adjacency matrix constructed as part of the model.
Let $\mathbf{A}$ be a $|\predicates{}| \times |\predicates{}|$ binary matrix
defined element-wise by stating that $\mathbf{A}[i][j] = 0$ if and only if, for
all $k = 0, \dots, \maxNumClauses{} - 1$, either
$\variable{heads}[k].\variable{predicate} \ne j$ or $ i \not\in
\{a.\variable{name} \mid a \in \variable{bodies}[k].\variable{values} \}$.

Given a partially-solved model with its predicate dependency graph, let us pick
an arbitrary path from $\mathsf{Q}$ to $\mathsf{P}$ (for some $\mathsf{P},
\mathsf{Q} \in \predicates{}$) that consists of determined edges that are
denoted by $1$ in $\mathbf{A}$ and potential/undetermined edges that are
denoted by $\{ 0, 1 \}$. Each such path characterises a \emph{(potential)
  dependency} $\mathsf{Q}$ for $\mathsf{P}$. We classify all such dependencies
into three classes depending on the number of undetermined edges on the path.
These classes are outlined in \cref{tbl:dependencies}, where $p$ represents the
dependency predicate $\mathsf{Q}$, and, in the case of $\AlmostDetermined{}$,
$(s, t) \in \predicates{}^2$ is the one undetermined edge on the path. For a
dependency $d$---regardless of its exact type---we will refer to its predicate
$p$ as $d.\mathsf{predicate}$. In describing the algorithms, we will use `$\_$'
to replace any of $p$, $s$, $t$ in situations where the name is unimportant.

\begin{algorithm}[t]
  \SetKwFunction{getDependencies}{deps}
  \SetKwFunction{isDetermined}{isDetermined}
  \SetKwData{predicate}{predicate}
  \KwData{predicates $p_1$, $p_2$}
  $D \gets \{ (d_1, d_2) \in \getDependencies{$p_1$, 1} \times
  \getDependencies{$p_2$, 1} \mid d_1.\predicate = d_2.\predicate \}$\;
  \lIf{$D = \emptyset$}{\Return{\textsc{true}}}
  \leIf{$\exists (\Determined \_, \Determined \_) \in
    D$}{\Return{\textsc{false}}}{\Return{\textsc{undefined}}}
  \caption{Entailment for independence}
  \label{alg:independence_entailment}
\end{algorithm}

Each entailment algorithm returns one out of three values:
\textsc{true} if the constraint is guaranteed to hold, \textsc{false} if the
constraint is violated, and \textsc{undefined} if whether the constraint will be
satisfied or not depends on the future decisions made by the solver.
\Cref{alg:independence_entailment} outlines a simple entailment algorithm for
the independence of two predicates $p_1$ and $p_2$. First, we separately
calculate all dependencies of $p_1$ and $p_2$ and look at the set $D$ of
dependencies that $p_1$ and $p_2$ have in common. If there are none, then the
predicates are clearly independent. If they have a dependency in common that is
already fully determined ($\Determined$) for both predicates, then they cannot be
independent. Otherwise, we return \textsc{undefined}.

\begin{algorithm}[t]
  \LinesNumbered
  \SetKwFunction{getDependencies}{deps}
  \SetKwFunction{fail}{fail}
  \SetKwFunction{removeValue}{removeValue}
  \SetKwData{predicate}{predicate}
  \SetKwData{source}{source}
  \SetKwData{target}{target}
  \KwData{predicates $p_1$, $p_2$; adjacency matrix $\mathbf{A}$}
  \For{$(d_1, d_2) \in \getDependencies{$p_1$, 0} \times \getDependencies{$p_2$,
      0}$ s.t. $d_1.\predicate = d_2.\predicate$}{
    \lIf{$d_1$ {\bf is} $\Determined(\_)$ {\bf and} $d_2$ {\bf is}
      $\Determined(\_)$}{\fail{}} \label{line:fail}
    \lIf{$\{ d_1, d_2 \} = \{ \Determined(\_), \AlmostDetermined(\_, s, t)
      \}$ \label{line:propagation_condition}}{$\mathbf{A}[s][t]$.\removeValue{$1$}}
  }
  \caption{Propagation for independence}
  \label{alg:independence_propagation}
\end{algorithm}

Propagation algorithms have two goals: causing a contradiction (failing) in
situations where the corresponding entailment algorithm would return
\textsc{false}, and eliminating values from domains of variables that are
guaranteed to cause a contradiction. \cref{alg:independence_propagation} does
the former on \cref{line:fail}. Furthermore, for any dependency shared between
predicates $p_1$ and $p_2$, if it is determined ($\Determined$) for one
predicate and almost determined ($\AlmostDetermined$) for another, then the edge
that prevents the $\AlmostDetermined$ from becoming a $\Determined$ cannot
exist---\cref{line:propagation_condition} handles this possibility.

\begin{algorithm}[t]
  \SetKwData{edgeExists}{edge}
  \SetKwData{predicate}{predicate}
  \SetKwData{source}{source}
  \SetKwData{target}{target}
  \SetKwData{all}{allDeps}
  \SetKwFunction{getDependencies}{deps}
  \SetKwProg{Fn}{Function}{:}{}
  \KwData{adjacency matrix $\mathbf{A}$}
  \Fn{\getDependencies{$p$, \all}} {
    $D \gets \{ \Determined(p) \}$\;
    \While{\logical{true}}{
      $D' \gets \emptyset$\;
      \For{$d \in D$ {\bf and} $q \in \predicates{}$}{
        $\edgeExists \gets \mathbf{A}[q][d.\predicate] = \{ 1 \}$\;
        \lIf{$\edgeExists$ {\bf and} $d$ {\bf is} $\Determined(\_)$}{$D' \gets
          D' \cup \{ \Determined(q) \}$}
        \lElseIf{$\edgeExists$ {\bf and} $d$ {\bf is} $\AlmostDetermined(\_, s,
          t)$}{$D' \gets D' \cup \{ \AlmostDetermined(q, s, t) \}$}
        \uElseIf{$|\mathbf{A}[q][d.\predicate]| > 1$ {\bf and} $d$~{\bf is}~$\Determined(r)$}{
          $D' \gets D' \cup \{ \AlmostDetermined(q, q, r) \}$\;
        }
        \lElseIf{$|\mathbf{A}[q][d.\predicate]| > 1$ {\bf and} \all}{$D' \gets
          D' \cup \{ \Undetermined(q) \}$}
      }
      \leIf{$D' = D$}{\Return{$D$}}{$D \gets D'$}
    }
  }
  \caption{Dependencies of a predicate}
  \label{alg:dependencies}
\end{algorithm}

The function $\getDependencies$ in \cref{alg:dependencies} calculates $D_p$ for
any predicate $p$. It has two versions: $\getDependencies(p, 1)$ returns all
dependencies, while $\getDependencies(p, 0)$ returns only determined and
almost-determined dependencies. It starts by establishing the predicate $p$
itself as a dependency and continues to add dependencies of dependencies until
the set $D$ stabilises. For each dependency $d \in D$, we look at the in-links
of $d$ in the predicate dependency graph. If the edge from some predicate $q$ to
$d.\mathsf{predicate}$ is fully determined and $d$ is determined, then $q$ is
another determined dependency of $p$. If the edge is determined but $d$ is
almost determined, then $q$ is an almost-determined dependency. The same outcome
applies if $d$ is fully determined but the edge is undetermined. Finally, if we
are interested in collecting all dependencies regardless of their status, then
$q$ is a dependency of $p$ as long as the edge from $q$ to
$d.\mathsf{predicate}$ is possible. Note that if there are multiple paths in the
dependency graph from $q$ to $p$, \cref{alg:dependencies} could include $q$ once
for each possible type ($\Determined$, $\Undetermined$, and
$\AlmostDetermined$), but
\cref{alg:independence_propagation,alg:independence_entailment} would still work
as intended.

\begin{figure}[t]
    \begin{minipage}{.49\textwidth}
      \centering
      \subfloat[The adjacency matrix of the graph. The boxed value is the
      decision variable that will be propagated by
      \cref{alg:independence_propagation}.\label{fig:dependencies_matrix}]{
        $\begin{blockarray}{lcccc}
          \begin{block}{l!{\quad}(cccc)}
            \mathsf{father} & 0 & 0 & 0 & 0 \\
            \mathsf{mother} & 1 & 0 & 0 & 0 \\
            \mathsf{parent} & 1 & \fbox{\{ 0, 1 \}} & \{ 0, 1 \} & \{ 0, 1 \} \\
            \mathsf{sibling} & 0 & 0 & 0 & 0 \\
          \end{block}
        \end{blockarray}$
      }
    \end{minipage}
    \hfill
    \begin{minipage}{.49\textwidth}
      \centering
      \subfloat[A drawing of the graph. Dashed edges are undetermined---they may
      or may not exist.\label{fig:dependencies2}]{
        \centering
        \makebox[\linewidth]{
          \begin{tikzpicture}
            \node[draw] (parent) at (0, 0.5) {$\mathsf{parent}$};
            \node[draw] (mother) at (0, -0.5) {$\mathsf{mother}$};
            \node[draw] (sibling) at (2, 0.5) {$\mathsf{sibling}$};
            \node[draw] (father) at (2, -0.5) {$\mathsf{father}$};
            \draw[-{Stealth}] (parent) edge (father);
            \draw[-{Stealth}] (mother) edge (father);
            \draw[-{Stealth},dashed,darkgray] (parent) edge (sibling);
            \draw[-{Stealth},dashed,darkgray] (parent) edge (mother);
          \end{tikzpicture}
        }
      }
    \end{minipage}
    \caption{The predicate dependency graph of \protect{\cref{example:independence}}}
    \label{fig:example}
  \end{figure}
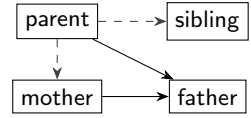

\begin{example} \label{example:independence}
  Consider this partially determined (fragment of a) program:
  \begin{align*}
    \Box(X, Y) &\gets \mathsf{parent}(X, Z) \land \mathsf{parent}(Y, Z),\\
    \mathsf{father}(X, Y) &\gets \mathsf{parent}(X, Y) \land \neg\mathsf{mother}(X, Y),
  \end{align*}
  where $\Box$ indicates an unknown predicate with domain
  \[
    D_\Box = \{ \mathsf{father}, \mathsf{mother}, \mathsf{parent},
    \mathsf{sibling} \}.
  \]
  The predicate dependency graph is pictured in \cref{fig:example}. Suppose we
  have a constraint that $\mathsf{mother}$ and $\mathsf{parent}$ must be
  independent. The lists of potential dependencies for both predicates are:
  \begin{align*}
    D_{\mathsf{mother}} &= \{ \Determined(\mathsf{mother}), \AlmostDetermined(\mathsf{parent}, \mathsf{parent}, \mathsf{mother}) \}, \\
    D_{\mathsf{parent}} &= \{ \Determined(\mathsf{parent}) \}.
  \end{align*}
  An entailment check at this stage would produce \textsc{undefined}, but
  propagation replaces the boxed value in \cref{fig:dependencies_matrix} with
  zero, eliminating the potential edge from $\mathsf{parent}$ to
  $\mathsf{mother}$. This also eliminates $\mathsf{mother}$ from $D_\Box$, and
  this is enough to make \cref{alg:independence_entailment} return
  \textsc{true}.
\end{example}

\section{Experimental Results} \label{sec:experiments}

We now present the results of two experiments: in \cref{sec:experiment1} we
examine the scalability of our constraint model with respect to its parameters
and in \cref{sec:experiment2} we demonstrate how the model can be used to
compare inference algorithms and describe their behaviour across a wide range of
programs. The experiments were run on a system with Intel Core i5-8250U
processor and \SI{8}{\giga\byte} of RAM. The constraint model was implemented in
Java~8 with Choco~4.10.2 \cite{choco}. All inference algorithms are implemented
in \textsc{ProbLog}~2.1.0.39 and were run using Python~3.8.2 with PySDD~0.2.10
and PyEDA~0.28.0. For both sets of experiments, we generate programs without
negative cycles and use a \SI{60}{\second} timeout.

\subsection{Empirical Performance of the Model} \label{sec:experiment1}

Along with constraints, variables, and their domains, two more design decisions
are needed to complete the model: heuristics and restarts. By trial and error,
the variable ordering heuristic was devised to eliminate sources of
\emph{thrashing}, i.e., situations where a contradiction is being `fixed' by
making changes that have no hope of fixing the contradiction. Thus, we partition
all decision variables into an ordered list of groups and require the values of
all variables from one group to be determined before moving to the next group.
Within each group, we use the `fail first' variable ordering heuristic. The
first group consists of all head predicates. Afterwards, we handle all remaining
decision variables from the first clause before proceeding to the next. The
decision variables within each clause are divided into
\begin{enumerate*}[(a)]
\item the $\variable{structure}$ array,
\item body predicates,
\item head arguments,
\item (if $|\variables{}| > 1$) the $\variable{intros}$ array,
\item body arguments.
\end{enumerate*}
For instance, in the clause from \cref{example:sibling}, all visible parts of
the clause would be decided in this order:
\[
  \overset{1}{\mathsf{sibling}}(\overset{3}{X}, \overset{3}{Y}) \gets
  \overset{2}{\mathsf{parent}}(\overset{4}{X}, \overset{4}{Z})
  \overset{2}{\land} \overset{2}{\mathsf{parent}}(\overset{4}{Y},
  \overset{4}{Z}).
\]
We also employ a geometric restart policy, restarting after $10, 10 \times 1.1,
10 \times 1.1^2, \dots$ contradictions.\footnote{Restarts help overcome early
  mistakes in the search process but can be disabled if one wants to find all
  solutions, in which case search is complete regardless of the variable
  ordering heuristic.} We ran \num{399360} experiments, investigating the
model's efficiency and gaining insight into what parameter values make the CSP
harder. For $|\predicates{}|$, $|\variables{}|$, $|\constants{}|$,
$\maxNumNodes{}$, and $\maxNumClauses{} - |\predicates{}|$ (i.e., the number of
clauses in addition to the mandatory $|\predicates{}|$ clauses), we assign all
combinations of 1, 2, 4, 8. $\maxArity{}$ is assigned to values 1--4. For each
$|\predicates{}|$, we also iterate over all possible numbers of independent
pairs of predicates, ranging from 0 up to $\binom{|\predicates{}|}{2}$. For each
combination of the above-mentioned parameters, we pick ten random ways to assign
arities to predicates (such that $\maxArity{}$ occurs at least once) and ten
random combinations of independent pairs.

The majority (\SI{97.7}{\percent}) of runs finished in under \SI{1}{\second},
while four instances timed out: all with $|\predicates{}| = \maxNumClauses{} -
|\predicates{}| = \maxNumNodes{} = 8$ and the remaining parameters all
different. This suggests that---regardless of parameter values---most of the
time a solution can be identified instantaneously while occasionally a series of
wrong decisions can lead the solver into a part of the search space with no
solutions.

In \cref{fig:impact}, we plot how the mean number of nodes in the binary search
tree grows as a function of each parameter (the plot for the median is very
similar). The growth of each curve suggests how the model scales with higher
values of the parameter. From this plot, it is clear that $\maxNumNodes{}$ is
the limiting factor. This is because some tree structures can be impossible to
fill with predicates without creating either a negative cycle or a forbidden
dependency, and such trees become more common as the number of nodes increases.
Likewise, a higher number of predicates complicates the situation as well.

\begin{figure}[t]
  \centering
  \begin{minipage}[t]{.49\textwidth}
    \centering
    \input{impact.tex}%
    \captionof{figure}{The mean number of nodes in the binary search tree for
      each value of each experimental parameter. Note that the horizontal axis
      is on a $\log_2$ scale.}
    \label{fig:impact}
  \end{minipage}
  \hfill
  \begin{minipage}[t]{.49\textwidth}
    \centering
    \input{bars.tex}%
    \captionof{figure}{Inference time for different values of $\maxArity{}$ and
      proportions of probabilistic facts that are probabilistic. The total
      number of facts is fixed at $10^5$.}
    \label{fig:bars}
  \end{minipage}
\end{figure}

\subsection{Experimental Comparison of Inference Algorithms} \label{sec:experiment2}

For this experiment, we consider clauses of two types: \emph{rules} are clauses
such that the head atom has at least one variable, and \emph{facts} are clauses
with empty bodies and no variables. We use our constraint model to generate the
rules according to the following parameter values: $|\predicates{}|,
|\variables{}|, \maxNumNodes{} \in \{ 2, 4, 8 \}$, $\maxArity{} \in \{ 1, 2, 3
\}$, $\maxNumClauses{} = |\predicates{}|$, $\constants{} = \emptyset$. These
values are (approximately) representative of many standard benchmarking
instances which often have 2--8 predicates of arity one or two, 0--8 rules, and
a larger database of facts \cite{DBLP:journals/tplp/FierensBRSGTJR15}. Just like
before, we explore all possible numbers of independent predicate pairs. We also
add a constraint that forbids empty bodies. For both rules and facts,
probabilities are uniformly sampled from $\{ 0.1, 0.2, \dots, 0.9 \}$.
Furthermore, all rules are probabilistic, while we vary the proportion of
probabilistic facts among \SI{25}{\percent}, \SI{50}{\percent}, and
\SI{75}{\percent}. For generating facts, we consider $|\constants{}| \in \{100,
200, 400 \}$ and vary the number of facts among $10^3$, $10^4$, and $10^5$ but
with one exception: the number of facts is not allowed to exceed
\SI{75}{\percent} of all possible facts with the given values of
$\predicates{}$, $\arities{}$, and $\constants{}$. Facts are generated using a
simple procedure that randomly selects a predicate, combines it with the right
number of constants, and checks whether the generated atom is already included
or not. We randomly select configurations from the description above and
generate ten programs with a complete restart of the constraint solver before
the generation of each program, including choosing different arities and
independent pairs. Finally, we set the query of each program to a random fact
not explicitly included in the program and consider six natively supported
algorithms and knowledge compilation techniques: binary decision diagrams (BDDs)
\cite{DBLP:journals/tc/Bryant86}, negation normal form (NNF), deterministic
decomposable NNF (d-DNNF) \cite{DBLP:journals/jair/DarwicheM02}, K-Best
\cite{DBLP:conf/ijcai/RaedtKT07}, and two encodings based on sentential decision
diagrams \cite{DBLP:conf/ijcai/Darwiche11}, one of which encodes the entire
program (SDDX), while the other one encodes only the part of the program
relevant to the query (SDD).\footnote{Forward SDDs (FSDDs) and forward BDDs
  (FBDDs) \cite{DBLP:conf/aaai/TsamouraGK20,DBLP:conf/ijcai/VlasselaerBKMR15}
  are omitted because the former uses too much memory and the implementation of
  the latter seems to be broken at the time of writing.}

\begin{figure}[t]
  \centering
  \input{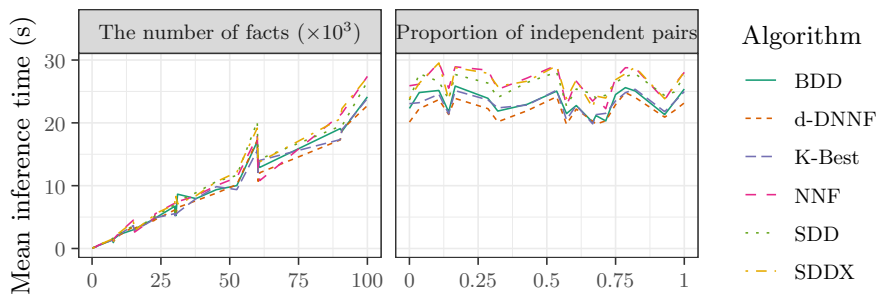}%
  \caption{Mean inference time for a range of \textsc{ProbLog} inference
    algorithms as a function of the total number of facts in the program and the
    proportion of independent pairs of predicates. For the second plot, the
    number of facts is fixed at $10^5$.}
  \label{fig:line_plots}
\end{figure}

Out of \num{11310} generated problem instances, about \SI{35}{\percent} were
discarded because one or more algorithms were not able to ground the instance
unambiguously. The first observation (pictured in \cref{fig:line_plots}) is that
the algorithms are remarkably similar, i.e., the differences in performance are
small and consistent across all parameter values (including parameters not shown
in the figure). Unsurprisingly, the most important predictor of inference time
is the number of facts. However, after fixing the number of facts to a constant
value, we can still observe that inference becomes harder with higher arity
predicates as well as when facts are mostly probabilistic (see \cref{fig:bars}).
Finally, according to \cref{fig:line_plots}, the independence structure of a
program does not affect inference time, i.e., state-of-the-art inference
algorithms---although they are supposed to
\cite{DBLP:conf/uai/FierensBTGR11}---do not exploit situations where separate
parts of a program can be handled independently.

\section{Conclusion}

We described a constraint model for generating both logic programs and
probabilistic logic programs. The model avoids unnecessary symmetries, is
reasonably efficient and supports additional constraints such as predicate
independence. Our experimental results provide the first comparison of inference
algorithms for probabilistic logic programming languages that generalises over
programs, i.e., is not restricted to just a few programs and data sets. While
the results did not reveal any significant differences among the algorithms,
they did reveal a shared weakness, i.e., the inability to ignore the part of a
program that is easily seen to be irrelevant to the given query.

Nonetheless, we would like to outline two directions for future work. First, the
experimental evaluation in \cref{sec:experiment1} revealed scalability issues,
particularly concerning the length/complexity of clauses. However, this
particular issue is likely to resolve itself if the format of a clause is
restricted to a conjunction of literals. Second, random instance generation
typically focuses on either realistic instances or sampling from a simple and
well-defined probability distribution. Our approach can be used to achieve the
former, but it is an open question how it could accommodate the latter.

\subsubsection*{Acknowledgments.}
Paulius was supported by the EPSRC Centre for Doctoral Training in Robotics
and Autonomous Systems, funded by the UK Engineering and Physical Sciences
Research Council (grant EP/S023208/1). Vaishak was supported by a Royal Society
University Research Fellowship.

\bibliographystyle{splncs04}
\bibliography{paper}

\newpage
\appendix
\section{Example Programs}

In this appendix, we provide examples of probabilistic logic programs generated
by various combinations of parameters. In all cases, we use
\[
  \{ 0.1, 0.2, \dots, 0.9, 1, 1, 1, 1, 1\}
\]
as the multiset of probabilities. Each clause is written on a separate line and
ends with a full stop. The head and the body of each clause are separated with
\texttt{:-} (instead of $\gets$). The probability of each clause is prepended to
the clause, using \texttt{::} as a separator. Probabilities equal to one and
empty bodies of clauses can be omitted. Conjunction, disjunction, and negation
are denoted by commas, semicolons, and `\texttt{\textbackslash+}', respectively.
Parentheses are used to demonstrate precedence, although many of them are
redundant. 

By setting $\predicates{} = [\texttt{p}]$, $\arities{} = [1]$, $\variables{} =
\{ \texttt{X} \}$, $\constants{} = \emptyset$, $\maxNumNodes{} = 4$, and
$\maxNumClauses{} = 1$, we get fifteen one-line programs, six of which are
without negative cycles (as highlighted below). Only the last program has no
cycles at all.

\begin{enumerate}[label=\arabic*.]
\item
\begin{verbatim}
0.5 :: p(X) :- (\+(p(X))), (p(X)).
\end{verbatim}
\item
\begin{verbatim}
0.8 :: p(X) :- (\+(p(X))); (p(X)).
\end{verbatim}
\hlitem
\begin{verbatim}
0.8 :: p(X) :- (p(X)); (p(X)).
\end{verbatim}
\hlitem
\begin{verbatim}
0.7 :: p(X) :- (p(X)), (p(X)).
\end{verbatim}
\item
\begin{verbatim}
0.6 :: p(X) :- (p(X)), (\+(p(X))).
\end{verbatim}
\item
\begin{verbatim}
p(X) :- (p(X)); (\+(p(X))).
\end{verbatim}
\hlitem
\begin{verbatim}
0.1 :: p(X) :- (p(X)); (p(X)); (p(X)).
\end{verbatim}
\hlitem
\begin{verbatim}
0.8 :: p(X) :- (p(X)), (p(X)), (p(X)).
\end{verbatim}
\item
\begin{verbatim}
p(X) :- \+(p(X)).
\end{verbatim}
\item
\begin{verbatim}
0.1 :: p(X) :- \+(\+(p(X))).
\end{verbatim}
\item
\begin{verbatim}
p(X) :- \+((p(X)); (p(X))).
\end{verbatim}
\item
\begin{verbatim}
0.4 :: p(X) :- \+((p(X)), (p(X))).
\end{verbatim}
\item
\begin{verbatim}
0.4 :: p(X) :- \+(\+(\+(p(X)))).
\end{verbatim}
\hlitem
\begin{verbatim}
0.7 :: p(X) :- p(X).
\end{verbatim}
\hlitem
\begin{verbatim}
p(X).
\end{verbatim}
\end{enumerate}

Note that:
\begin{itemize}
\item A program such as Program~14, because of its cyclic definition, defines a
  predicate that has probability zero across all constants. This can more easily
  be seen as solving equation $0.7x = x$.
\item Programs 10 and 14 are not equivalent (i.e., double negation does not
  cancel out) because Program 10 has a negative cycle and is thus considered to
  be ill-defined.
\end{itemize}

To demonstrate variable symmetry reduction in action, we set $\predicates{} =
[\texttt{p}]$, $\arities{} = [3]$, $\variables{} = \{\texttt{X}, \texttt{Y},
\texttt{Z} \}$, $\constants{} = \emptyset$, $\maxNumNodes{} = 1$,
$\maxNumClauses{} = 1$, and forbid all cycles. This gives us the following five
programs:

\begin{itemize}
\item
\begin{verbatim}
0.8 :: p(Z, Z, Z).
\end{verbatim}
\item
\begin{verbatim}
p(Y, Y, Z).
\end{verbatim}
\item
\begin{verbatim}
p(Y, Z, Z).
\end{verbatim}
\item
\begin{verbatim}
p(Y, Z, Y).
\end{verbatim}
\item
\begin{verbatim}
0.1 :: p(X, Y, Z).
\end{verbatim}
\end{itemize}

This is one of many possible programs with $\predicates{} = [\texttt{p},
\texttt{q}, \texttt{r}]$, $\arities{} = [1, 2, 3]$, $\variables{} =
\{\texttt{X}, \texttt{Y}, \texttt{Z} \}$, $\constants{} = \{ \texttt{a},
\texttt{b}, \texttt{c} \}$, $\maxNumNodes{} = 5$, $\maxNumClauses{} = 5$, and
without negative cycles:

\begin{verbatim}
p(b) :- \+((q(a, b)), (q(X, Y)), (q(Z, X))).
0.4 :: q(X, X) :- \+(r(Y, Z, a)).
q(X, a) :- r(Y, Y, Z).
q(X, a) :- r(Y, b, Z).
r(Y, b, Z).
\end{verbatim}

Finally, we set
$\predicates{} = [\texttt{p}, \texttt{q}, \texttt{r}]$, $\arities{} = [1, 1,
1]$, $\variables{} = \emptyset$, $\constants{} = \{ \texttt{a} \}$,
$\maxNumNodes{} = 3$, $\maxNumClauses{} = 3$, forbid negative cycles, and
constrain predicates \texttt{p} and \texttt{q} to be independent. The resulting
search space contains thousands of programs such as:

\begin{itemize}
\item
\begin{verbatim}
0.5 :: p(a) :- (p(a)); (p(a)).
0.2 :: q(a) :- (q(a)), (q(a)).
0.4 :: r(a) :- \+(q(a)).
\end{verbatim}
\item
\begin{verbatim}
p(a) :- p(a).
0.5 :: q(a) :- (r(a)); (q(a)).
r(a) :- (r(a)); (r(a)).
\end{verbatim}
\item
\begin{verbatim}
p(a) :- (p(a)); (p(a)).
0.6 :: q(a) :- q(a).
0.7 :: r(a) :- \+(q(a)).
\end{verbatim}
\end{itemize}

\end{document}